\documentclass[conference]{IEEEtran}
\IEEEoverridecommandlockouts
\usepackage{cite}
\usepackage{algorithm}
\usepackage{array}
\usepackage[caption=false,font=normalsize,labelfont=sf,textfont=sf]{subfig}
\usepackage{adjustbox}
\usepackage{makecell}
\usepackage{textcomp}
\usepackage{stfloats}
\usepackage{url}
\usepackage{amsmath,amssymb,amsfonts}
\usepackage{verbatim}  
\usepackage{graphicx}
\usepackage{multirow}
\DeclareMathOperator*{\argmin}{argmin}
\usepackage{algorithmic}
\usepackage{xcolor}
\def\BibTeX{{\rm B\kern-.05em{\sc i\kern-.025em b}\kern-.08em
    T\kern-.1667em\lower.7ex\hbox{E}\kern-.125emX}}
\begin{document}

\title{AE-DENet: Enhancement for Deep Learning-based Channel Estimation in OFDM Systems

\thanks{*Corresponding author. This project is funded by the National Key Research and Development Program of China (2023YFC3806003, 2023YFC3806001).}
}
\author{\IEEEauthorblockN{ Ephrem Fola, Yang Luo, Chunbo Luo*}
\IEEEauthorblockA{\textit{School of Information and Communication Engineering } \\
\textit{University of Electronic Science and Technology of China, Chengdu 611731, China }\\
Email: ephrem.b@std.uestc.edu.cn, \{luoyang, c.luo\}@uestc.edu.cn}

}

\maketitle

\begin{abstract}
Deep learning (DL)-based methods have demonstrated remarkable achievements in addressing orthogonal frequency division multiplexing (OFDM) channel estimation challenges. However, existing DL-based methods mainly rely on separate real and imaginary inputs while ignoring the inherent correlation between the two streams, such as amplitude and phase information that are fundamental in communication signal processing. This paper proposes AE-DENet, a novel autoencoder(AE)-based data enhancement network to improve the performance of existing DL-based channel estimation methods. AE-DENet focuses on enriching the classic least square (LS) estimation input commonly used in DL-based methods by employing a learning-based data enhancement method, which extracts interaction features from the real and imaginary components and fuses them with the original real/imaginary streams to generate an enhanced input for better channel inference. Experimental findings in terms of the mean square error (MSE) results demonstrate that the proposed method enhances the performance of all state-of-the-art DL-based channel estimators with negligible added complexity. Furthermore, the proposed approach is shown to be robust to channel variations and high user mobility.
\end{abstract}


\section{Background and motivations}
\IEEEPARstart{O}{rthogonal} Frequency Division Multiplexing (OFDM), is a key modulation technique in the fourth-generation (4G) long-term evolution (LTE), and fifth-generation new radio (5G NR) cellular network standards and will continue to be a key technology in the future sixth-generation (6G) systems \cite{Ozpoyraz}. Accurate channel estimation is crucial in OFDM systems to meet the extremely high data rate and low latency requirements. Traditional channel estimation techniques typically include two main types of estimators: the least square (LS) estimator which is computationally efficient but yields comparatively poor performance due to its sensitivity to channel noise, and the minimum mean squared error (MMSE) estimator which gives optimal estimation results using the prior knowledge of channel statistics but has high computational complexity. Further, obtaining prior channel knowledge is also difficult in some scenarios \cite{9128890}.

Deep learning (DL)-based channel estimation methods have recently been presented to overcome the limitations of traditional estimation methods\cite{8052521,8715338,8509622}. Particularly, model-assisted DL-based channel estimation approaches initialize the DL models using classical model-based channel estimation methods such as LS, reducing the demand for large training data. 
Typical examples include the ChanEstNet\cite{8761312} that uses initial LS estimation for the convolutional neural network (CNN) and long short-term memory (LSTM)-based channel estimation, and ChannelNet\cite{8640815} that models the channel time-frequency response as a 2-dimensional low-resolution image and uses the super-resolution CNN (SRCNN) and denoising CNN (DnCNN) to obtain the estimated channel. Excellent work has been carried out to further address more practical issues such as millimeter wave (mmWave) communications\cite{9815290}, IEEE 802.11p systems \cite{9120030}, and fast time-varying channels \cite{8933411} et al. 
The aforementioned model-assisted DL-based channel estimation approaches adopt separate real and imaginary values of the complex LS channel estimation as their inputs, while ignore the inherent correlation between the two streams which contain essential information such as amplitude and phase, as purposely designed. This loss of valuable information about the estimated channel inevitably impacts the estimation performance, particularly at lower signal-to-noise ratio (SNR) regions where the channel noise is usually severe.

Autoencoders (AEs) have been used in different applications in communication systems to capture the essential features of the input data through low-dimensional representations \cite{10052744,10051858,9166541}. Particularly, in \cite{10051858} a variational AE (VAE), which uses a continuous latent space to learn the distribution of the underlying input data, is proposed to approximate the MMSE channel estimator. In \cite{9166541}, an AE-based feature selection method has been applied to find the near-optimal pilot pattern for DL-based channel estimation in OFDM systems. In\cite{chen}, AE has been used to predict the spatial and temporal distribution features of continuous signals by extracting sequence features in intelligent reflecting surface-assisted mmWave communications.

Inspired by the success of AEs in feature extraction tasks, in this paper, we propose an AE-based data enhancement neural network, AE-DENet, to enhance DL-based channel estimation methods. The proposed AE-DENet incorporates an AE built with fully connected (FC) layer to extract the interactive features of the joint real/imaginary data of the initial LS estimation. These interactive features are then concatenated with the original real/imaginary streams to create an enhanced dataset, which can be used to train DL-based channel estimation models for fine channel estimation. The contributions of this paper are summarized as follows:

 \begin{enumerate}
 \item We propose AE-DENet, to incorporate the classic LS method to generate coarse channel estimation based on expert knowledge and a data-driven AE-based module to enhance the LS outputs with cross real/imaginary features.
 
 \item We verify the effectiveness of the proposed AE-DENet method for a range of state-of-the-art DL-based channel estimation methods in extensive simulation experiments on two typical 3$^{\text{rd}}$ generation partnership project (3GPP) channel models: Extended Pedestrian A (EPA) and Extended Vehicular A (EVA), which confirm its robust performance on varying channel complexities with a SNR gain about 3 dB and 2 dB in EPA and EVA channels, respectively.

 \item We further demonstrate that the input dimensions of the DL-based channel estimation models can be adaptively adjusted to accommodate the enhanced initial channel estimation, with minimal added complexity.
\end{enumerate}

The rest of this paper is organized as follows. Section \ref{section2} presents an overview of the system model and traditional channel estimators. In Section \ref{section3}, the proposed model-assisted channel estimation method is presented. Section \ref{section4} describes the simulation results to verify the effectiveness of the proposed approach. Conclusion is given in Section \ref{section5}.

\section{System model and traditional channel estimation methods} \label{section2}

At the receiver side of an OFDM system, the signal transmitted through a frequency selective multipath fading channel can be obtained by performing the discrete Fourier transform (DFT) on the received signal, which is given by:
\begin{equation}
    \boldsymbol{Y}=\boldsymbol{H} \circ \boldsymbol{X} + \boldsymbol{W},
\end{equation}
where $\boldsymbol{X}$, $\boldsymbol{Y}\in\mathbb{C}^{N_f\times N_s}$ are the DFTs of the transmitted and received signals, respectively.
$\boldsymbol{H} \in \mathbb{C}^{N_f\times N_s}$ is the frequency domain response of the channel while  $N_f$ and $N_s$ represent the deployed number of subcarriers and OFDM symbols, respectively. The operator $\circ$ is the Hadamard product (element-wise product) while $\boldsymbol{W} \in\mathbb{C}^{N_f\times N_s}$ is the additive white Gaussian noise (AWGN) with zero mean and variance $\sigma^2$.
Assuming pilot-based channel estimation where the pilot symbols are sparsely and evenly placed in the time-frequency grid, the frequency domain representation of the received pilot signals, $\boldsymbol{Y_p} \in \mathbb{C}^{N_{pf}\times N_{ps}}$, can be given by:
\begin{equation}
    \boldsymbol{Y_p}=\boldsymbol{H_p} \circ \boldsymbol{X_p} + \boldsymbol{W_p},
\end{equation}  
where $\boldsymbol{X_p}\in\mathbb{C}^{N_{pf}\times N_{ps}}$ denotes the transmitted pilot signals, $\boldsymbol{H_p} \in \mathbb{C}^{N_{pf}\times N_{ps}}$ is the frequency domain response of the channel at the pilot locations. $N_{pf}$ and $N_{ps}$ represent the number of pilot subcarriers placed along the subcarrier axis and the pilot symbols placed along the OFDM symbol axis in the time-frequency grid, respectively. $\boldsymbol{W_p}$ denotes the noise samples at the pilot positions. 

The LS channel estimation at the pilot positions can be obtained by solving the minimization problem given by:
\begin{equation}
\hat{\boldsymbol{H}}_{LS,p}=\argmin_{\boldsymbol{H_p}}{\left|\left| \boldsymbol{Y_p}- \boldsymbol{H_p} \circ \boldsymbol{X_p} \right|\right|_2^2}=\boldsymbol{Y_p}\boldsymbol{X_p}^{-1}.
\end{equation}
The LS estimate of the complete channel frame, $\hat{\boldsymbol{H}}_{LS} \in \mathbb{C}^ {N_f\times N_s}$ is obtained by applying a bilinear interpolation in both time and frequency domains.
The LS estimate is computationally simple but yields poor estimation results due to its sensitivity to the presence of noise in the channel. On the other hand, the MMSE channel estimation provides optimal results by utilizing prior channel knowledge. The frequency domain linear MMSE can be given by:
\begin{equation}
 \hat{\boldsymbol{H}}_{MMSE,p}=\boldsymbol{R_{hh_p}}\left(\boldsymbol{R_{h_ph_p}}+\frac{\sigma^2_N}{\sigma^2_X}\boldsymbol{I}\right)^{-1}\hat{\boldsymbol{H}}_{LS,p},   
\end{equation}
where $\boldsymbol{R_{hh_p}}=\mathbb{E}\{\boldsymbol{h}\boldsymbol{h_p}^{\boldsymbol{H}}\}$ and $\boldsymbol{R_{h_ph_p}}=\mathbb{E}\{\boldsymbol{h_p}\boldsymbol{h_p}^{\boldsymbol{H}}\}$ are the channel cross-correlation and autocorrelation matrices, respectively. $\boldsymbol{h_p}$ and $\boldsymbol{h}$ are the channel responses at pilot subcarriers and all subcarriers, respectively. $\boldsymbol{I}$ is the identity matrix while $\sigma^2_N$ and $\sigma^2_X$ represent the variances of AWGN noise and the average power of the transmitted signal, respectively. The complete MMSE estimate is obtained by applying a linear interpolation in the time domain. The MMSE estimation has higher computational complexity and requires prior channel knowledge which is not available in practice.

\section{Enhancement method for DL-based channel estimation: AE-DENet} \label{section3}

In terms of a mathematical expression, the received complex signal consists of the real, $\Re(\boldsymbol{Y})$, and imaginary, $\Im(\boldsymbol{Y})$, components which convey essential information about the received signal such as amplitude and phase, which can be expressed as
\begin{equation}
\textbf{y}_a=\sqrt{[\Re(\boldsymbol{Y})]^2 + [\Im(\boldsymbol{Y})]^2}    
, \textbf{y}_\theta = \arctan \{\frac{\Im(\boldsymbol{Y})}{\Re(\boldsymbol{Y})}\},
\end{equation}
where $\textbf{y}_a$ and $\textbf{y}_\theta$ represent the amplitude and phase of the received signal, respectively. 
Most DL-based channel estimation models ignore the correlation between the real and imaginary streams of the complex LS channel estimation and treat them as two separate real-valued channels, affecting the estimation performance since the maximum information embedded in the complex representation is not exploited.
This paper aims to address the aforementioned problem by proposing AE-DENet, a learning-based data enhancement model. The proposed AE-DENet extracts interaction features from the real and imaginary parts of the complex LS estimate and combines them with the original real and imaginary components to enhance the input and improve channel estimation performance.

\subsection{AE-DENet architecture }

 \begin{figure}[h]
\centering
\includegraphics[,width=1\linewidth]{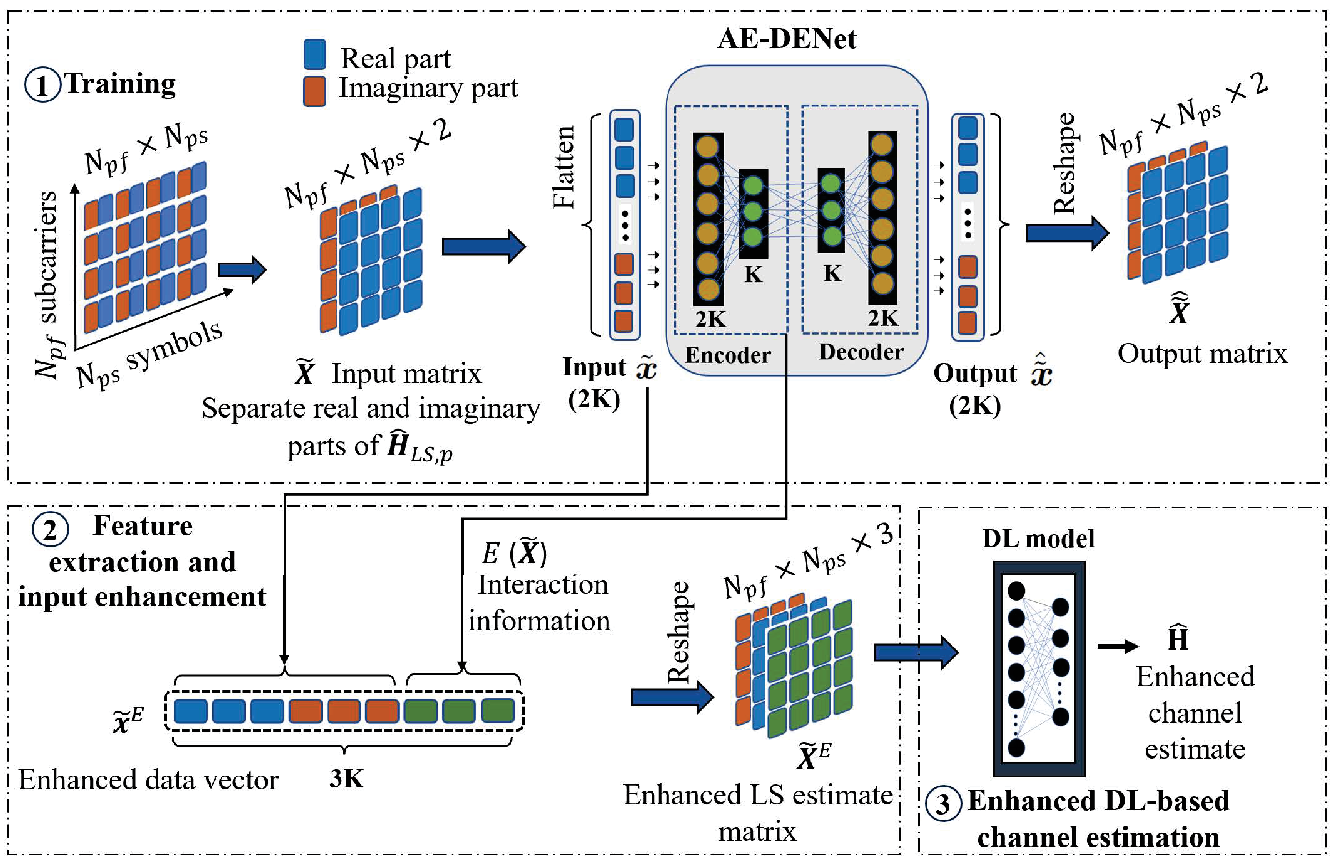}
\caption{The proposed AE-DENet model and the enhanced channel estimation approach.}
\label{fig:AE-DENet MODEL}
\end{figure}

The proposed AE-based data enhancement network, AE-DENet and the enhanced channel estimation approach are depicted in Fig.\ref{fig:AE-DENet MODEL}. AE-DENet is built with a FC network to extract the interactive features of real and imaginary streams of the LS estimation input, and generate enhanced LS estimation data for better channel inference. The proposed AE-DENet comprises an AE network with an encoder part to compress the LS input estimate into low-dimensional representation, and a decoder part to reconstruct the input. The AE employs an unsupervised learning technique that does not require explicit labels to train on.

In the proposed approach, we first compute the complex LS estimation matrix at the pilot positions $\hat{\boldsymbol{H}}_{LS,p} \in \mathbb{C}^ {N_{pf}\times N_{ps}}$ which will be either directly used or interpolated to complete channel frame ($\hat{\boldsymbol{H}}_{LS} \in \mathbb{C}^ {N_{f}\times N_{s}}$) depending on the input dimension of the particular DL-based channel estimation model. The complex LS estimation matrix is then split to separate real and imaginary parts to obtain a real-valued input matrix, $\tilde{\boldsymbol{X}}$, of dimension  $N_{pf}\times N_{ps}\times 2$ for $\hat{\boldsymbol{H}}_{LS,p}$ input or $N_{f}\times N_{s}\times 2$ for $\hat{\boldsymbol{H}}_{LS}$ input.
The input matrix, $\tilde{\boldsymbol{X}}$, is received by the AE and flattened to input vector $\tilde{\boldsymbol{x}}\in \mathbb{R}^ {2K}$, where $K = N_{pf} N_{ps}$ or $K = N_{f} N_{s}$. 
To accommodate the specific input dimensions of the representative DL-based channel estimation methods used in this paper, two kinds of AE-DENet configurations are considered:
$K = N_{pf} N_{ps}$ for $\hat{\boldsymbol{H}}_{LS,p}$ input and $K = N_{f} N_{s}$ for $\hat{\boldsymbol{H}}_{LS}$ input. 
In the training stage, as shown in Fig.\ref{fig:AE-DENet MODEL}, the input data will be processed by the FC layers of the encoder and decoder to give an output vector $\hat{\tilde{\boldsymbol{x}}}\in \mathbb{R}^ {2K}$, having the same dimension as $\tilde{\boldsymbol{x}}$. Following, the output vector $\hat{\tilde{\boldsymbol{x}}}$ is reshaped to the matrix $\hat{\tilde{\boldsymbol{X}}}$ to reconstruct the corresponding input matrix ($\hat{\boldsymbol{H}}_{LS,p}$ or $\hat{\boldsymbol{H}}_{LS}$). In the AE network, the first and second layers of the encoder part consist of $2K$ and $K$ nodes while the decoder's first and second layers have $K$ and $2K$ nodes, respectively.
The FC layers in the AE model facilitate feature interaction through full connectivity between each node in one layer with every node in the subsequent layer which plays a crucial role in learning the underlying structure of the input data. By associating the real and imaginary features with different weights of the nodes in the FC layers, the model can create interactive features that contain valuable information about the original complex LS estimation. For our proposed data enhancement method, we extract the interaction features from the encoder output with a dimension of $K$.
The AE-DENet contains a total of 4 FC layers each of them followed by a rectified linear unit (ReLU) activation function ($f_{ReLU}= max(0, x)$), except the last layer. Each of the encoder and the decoder parts contain 2 FC layers.
Given the set of all network parameters of the AE-DENet $\theta_{AE}$ and the input vector $\tilde{\boldsymbol{x}} \in \mathbb{R}^{2K} $, the recovered output matrix can be expressed as:
\begin{equation}
\hat{\tilde{\boldsymbol{X}}} =f_{AE} (\theta_{AE};\tilde{\boldsymbol{x}}), 
\end{equation}
where $f_{AE}$ is the AE-DENet's network function. The AE-DENet is trained with the objective of minimizing the mean squared error (MSE) loss, $L_{AE}$, between the input and the output channel of each resource element in the time-frequency grid, which is given by:
\begin{equation}
   L_{AE}=\frac{1}{N_f N_s}\sum^{N_f}_{k=1}\sum^{N_s}_{i=1}\left|\left| \hat{\tilde{\boldsymbol{X}}}_{ki}- {\tilde{\boldsymbol{X}}_{ki}} \right|\right|_2^2,
\end{equation}
where $\tilde{\boldsymbol{X}}_{ki}$ and $\hat{\tilde{\boldsymbol{X}}}_{ki}$ are the input channel and the reconstructed channel at the $k^{th}$ subcarrier and $i^{th}$ OFDM symbol, respectively.
After training the AE-DENet, we perform the feature extraction and input enhancement. In this stage, we feed the whole data sample to the trained AE-DENet to extract the interaction feature $E(\tilde{\boldsymbol{X}})$, with dimension $K$ from the encoder output, and then concatenate it with the original real and imaginary features to produce the enhanced data vector, $\tilde{\textbf{\textit{x}}}^E$, with dimension $3K$. This concatenated vector is finally reshaped to give the enhanced data matrix 
of the LS estimation, $\tilde{\boldsymbol{X}}^E$, with dimension $N_{pf}\times N_{ps}\times 3$ for $\hat{\boldsymbol{H}}_{LS,p}$ input or with dimension $N_{f}\times N_{s}\times 3$ for $\hat{\boldsymbol{H}}_{LS}$ input.

\subsection{Channel Estimation with AE-DENet Enhancement Method}

In the context of channel estimation, the AE-DENet model enhances the LS estimation by fusing the extracted interaction features with the original real and imaginary streams, forming an enriched input for improved channel inference. By incorporating AE-DENet for enhanced input data pre-processing, the estimation accuracy of existing DL-based channel estimation models can be improved.

To demonstrate the effectiveness of the proposed data enhancement method, several benchmark DL-based channel estimation models have been selected including CNN-based methods, SRCNN\cite{8640815} and ChannelNet\cite{8640815}, and the residual learning-based methods, ReEsNet\cite{li2019deep} and Interpolation-ResNet\cite{luan2021low}. For further performance comparison, we employ additional ReEsNet-12F and Interpolation-ResNet-12F models where the filter size of the convolutional layers of the corresponding ReEsNet and Interpolation-ResNet models is reduced from 16 to 12. Here, it should be noted that the input for both CNN-based models is the complete channel matrix, $N_{f}\times N_{s}\times 2$, which is obtained from the initial LS estimate at the pilot locations, $N_{pf}\times N_{ps}\times 2$, using linear interpolation. In contrast, the residual learning-based methods utilize the LS estimate at the pilot locations, which has a lower input dimension. In order to accommodate the AE-DENet's enhanced data dimension, the input dimension of the benchmark SRCNN and ChannelNet models is adjusted from $N_{f}\times N_{s}\times 2$  to $N_{f}\times N_{s}\times 3$ whereas the input dimension of the benchmark residual learning-based methods is adjusted from  $N_{pf}\times N_{ps}\times 2$ to $N_{pf}\times N_{ps}\times 3$.
We evaluate the performance of each benchmark DL-based channel estimation model with and without AE-DENet enhancement method by minimizing the MSE loss, $L_{CE}$, between the estimated channel and the actual channel of each resource element, which is given by: 
\begin{equation}
    L_{CE}=\frac{1}{N_f N_s}\sum^{N_f}_{k=1}\sum^{N_s}_{i=1}\left|\left| \hat{\boldsymbol{H}}_{ki}- \boldsymbol{H}_{ki} \right|\right|_2^2,
\end{equation}
where $\boldsymbol{H}_{ki}$  and $\hat{\boldsymbol{H}}_{ki}$ are the perfect channel, and the estimated channel at the $k^{th}$ subcarrier and $i^{th}$ OFDM symbol, respectively.

\section{Experimental results}\label{section4}

\subsection{Experimental setup}
We consider an OFDM system with $N_f=72$ subcarriers, equivalent to 6 5G resource blocks, and $N_s=14$ OFDM symbols. The carrier frequency is 2.1 GHz and the subcarrier spacing is 15 kHz. A cyclic prefix (CP) length of 16 OFDM samples and Quadrature Phase Shift Keying (QPSK) modulation are used.
For the pilot allocations in each frame, the $1^{\text{st}}$ and $13^{\text{th}}$ OFDM symbols are reserved for pilot symbols ($i=1, 13$). The pilot subcarriers of the $1^{\text{st}}$ pilot symbol are placed starting from the $1^{\text{st}}$ subcarrier index ($k=1$) with an interval of 3 subcarriers.
For the $2^{\text{nd}}$ pilot symbol, the pilot subcarriers are placed starting from the $2^{\text{nd}}$ subcarrier index ($k=2$) with the same interval of 3 subcarriers. Thus, a total of 48 pilots are used in each transmitted frame. 
A random mobile speed in the range of $0$ $km/h$ to $50$ $km/h$ is selected, corresponding to the maximum Doppler shift from 0 Hz
 to 97 Hz, respectively.

 \begin{table*}[!ht]  
  \caption{DL-based channel estimation models training specifications.}
   \centering
   \scalebox{0.9}{
    \begin{tabular}{c c c c c c}
    \hline 
\textbf{Parameter}  & \textbf{SRCNN} & \textbf{ChannelNet}& \textbf{ReEsNet/ ReEsNet-12F} & 
\makecell{\textbf{Interpolation-ResNet}/ \\ \textbf{Interpolation-ResNet-12F}}

        \\   
        \hline
    Number of Epochs  & 100 & 50 & 50 & 50 
     \\ 
    Initial learning rate (LR)  & 0.001 & 0.001 & 0.001 & 0.001 
        \\
    LR drop period  & None & None & every 20 epochs & every 20 epochs
         \\
    LR drop factor  & None & None & 0.5 & 0.5
    \\ 
   Minibatch size  & 128 & 128 & 128 & 128
      \\
  Optimizer  & ADAM & ADAM & ADAM & ADAM
       \\
 
 \hline
    \end{tabular}}
    \label{tab:Training specs}
\end{table*}

The proposed approach is evaluated with the widely used 3GPP channel models: EPA and EVA channel models, representing low and medium power delay profiles, respectively. 
The training dataset for both channel models is generated using MATLAB for SNR ranges between 0 dB to 20 dB (with an interval of 5 dB). For each SNR point, 5,000 channel realizations are generated. From a total of 25,000 data samples, 80\% is allocated for training while 20\% is used for validation. To enhance robustness, each DL model is trained with samples from a combination of all SNR ranges. In the testing stage, we generate a new dataset for both channel models using the same parameter settings as in the training stage while the SNR range is extended from 0 dB to 25 dB (a total of 30,000 data samples) to evaluate the generalization performance of the proposed method with unseen data samples. The MSE over the SNR and Doppler shift values are computed to compare the performances of the benchmark DL-based channel estimation models with those of their enhanced versions. The benchmark methods are SRCNN, ChannelNet, ReEsNet, ReEsNet-12F, Interpolation-ResNet and Interpolation-ResNet-12F while the corresponding modified models with enhanced input are named SRCNN-Enhanced, ChannelNet-Enhanced, ReEsNet-Enhanced, ReEsNet-12F-Enhanced, Interpolation-ResNet-Enhanced and Interpolation-ResNet-12F-Enhanced, respectively. Model training specifications for DL-based channel estimation models are presented in Table \ref{tab:Training specs}. The proposed AE-DENet is trained with the same training parameters as the SRCNN. All DL models are trained on a single NVIDIA GeForce GTX 1050Ti GPU.

\subsection{MSE versus SNR results }

\begin{figure}[ht!]
\centering
\includegraphics[,width=0.75\linewidth]{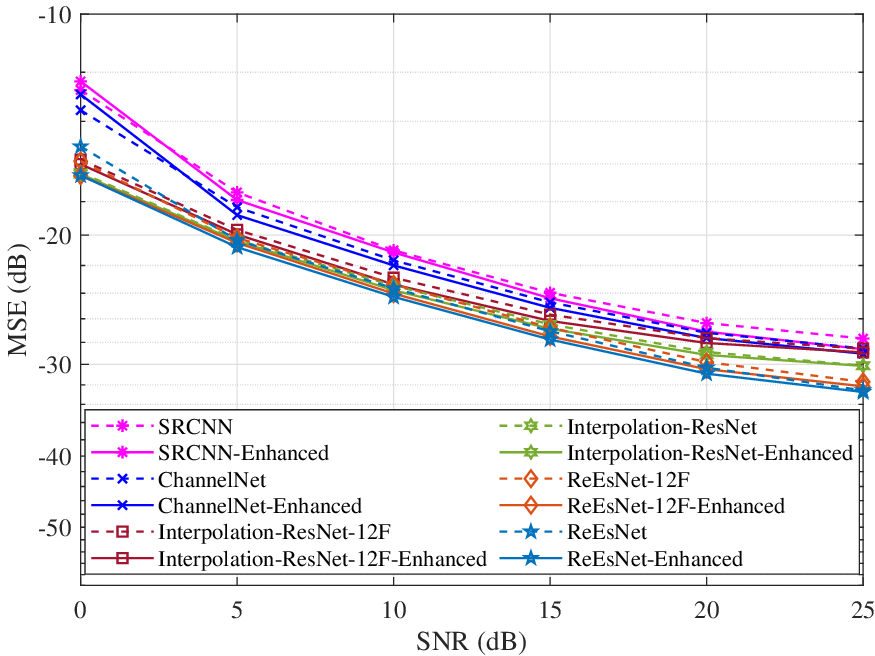}
\caption{MSE versus SNR results comparison of the enhanced and original DL-based channel estimation models for EPA channel model.}
\label{fig:MSE_SNR_EPA}
\end{figure}

\begin{figure}[h]
\centering
\includegraphics[,width=0.75\linewidth]{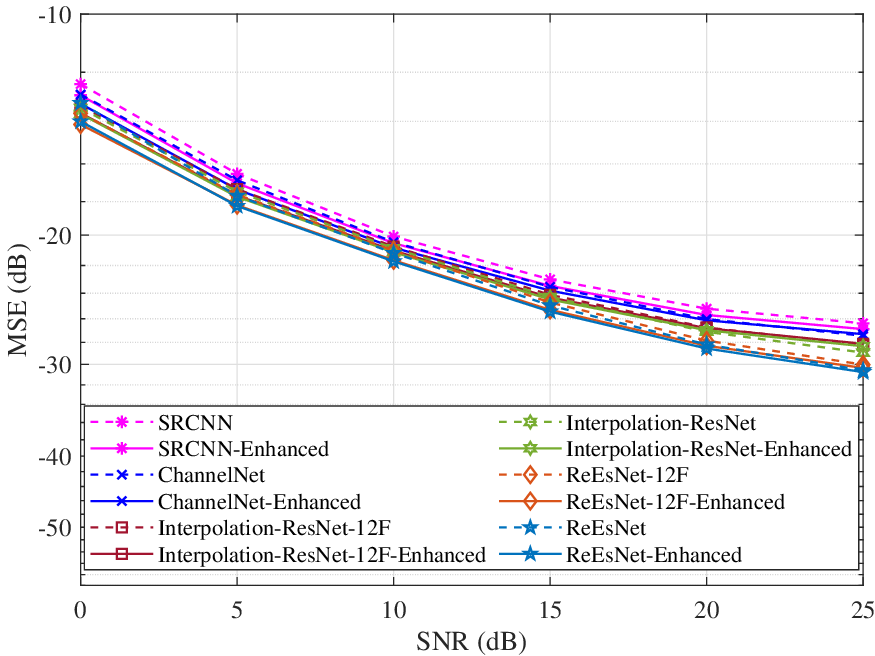}
\caption{MSE versus SNR results comparison of the enhanced and original DL-based channel estimation models for EVA channel model.}
\label{fig:MSE_SNR_EVA}
\end{figure}

In this section, we evaluate the MSE performances over the SNR range of each benchmark and enhanced models for the EPA and EVA channel models.
The result for the EPA channel in Fig.\ref{fig:MSE_SNR_EPA} shows that the enhanced models outperform the corresponding original models in most of the SNR ranges, with the best performance obtained for ReEsNet-Enhanced (MSE value of -32.72 dB at the maximum SNR value of 25 dB). At the high SNR ranges, a SNR gain of 3 dB is achieved for the SRCNN-Enhanced over the benchmark SRCNN. 
In Fig.\ref{fig:MSE_SNR_EVA}, the MSE results for the EVA channel show that, the performance of all methods is slightly lower than that of the EPA channel due to the relatively higher delay spread in the EVA channel environment. Nevertheless, the performance of the enhanced DL-based models is still better than the corresponding benchmark models in most of the SNR ranges. Particularly, the SRCNN-Enhanced has a 2 dB gain over the original SRCNN at the higher SNR values.

The challenge with the upscaling method (bilinear interpolation) used in both Interpolation-ResNet-Enhanced and Interpolation-ResNet-12F-Enhanced stems from its non-learnable nature. This limitation hampers the enhanced models’ ability to effectively leverage the refined input, which is drawn solely from pilot locations, to surpass the performance of the original models.
The MSE performance comparison of the original and the corresponding enhanced methods at lower SNR (5 dB) and higher SNR (20 dB) values is given in Table \ref{tab:MSE_comparison}. Notably, the enhanced methods exhibit better inference capability and are robust to channel variations (for EPA and EVA channel models) with higher estimation performance over the benchmark methods at the low and high SNR ranges.

 \begin{table*}[!ht]  
  \caption{MSE performance comparison of the original and enhanced models on EPA and EVA channel models.}
   \centering
   \scalebox{0.9}{
    \begin{tabular}{c c c c c c c c c}
    \hline 
 \makecell{\textbf{Channel}\\ \textbf{model}}   &  
 \makecell{\textbf{SNR}\\ \textbf{value}} &  &
\makecell{\textbf{SRCNN}}  & 
\makecell{\textbf{ChannelNet}} & 
\makecell{\textbf{ReEsNet}} & 
\makecell{\textbf{ReEsNet-12F}} & 
\makecell{\textbf{Interpolation-ResNet}} &
\makecell{\textbf{Interpolation-ResNet-12F}}
        \\   
        \hline
\multirow{4}{1cm}{\centering EPA} & \multirow{2}{1cm}{\centering 5 dB} &Original& -17.50 & -18.33 & -20.32 & -20.18 & -20.32 & -19.67 
     \\ 
       & &Enhanced& \textbf{-17.90} & \textbf{-18.79} & \textbf{-20.76} & \textbf{-20.51} & \textbf{-20.36} &\textbf{-19.96} 
        \\ \cline{2-9}
&\multirow{2}{1cm}{\centering 20 dB}  &Original& -26.38 & -27.21 & -30.32 & -29.59 & -28.86 & -27.70
         \\
       & &Enhanced& \textbf{-26.99} & \textbf{-27.70} & \textbf{-30.86} & \textbf{-30.46} & \textbf{-29.21} & \textbf{-27.96}
    \\ 
    \hline
    
\multirow{4}{1cm}{\centering EVA} & \multirow{2}{1cm}{\centering 5 dB}  &Original& -16.50 & -16.84 & -17.67 & -17.62 & -17.45 & -17.30 
     \\ 
       & &Enhanced& \textbf{-17.01} & \textbf{-17.33} & \textbf{-18.24} & \textbf{-18.21} & \textbf{-17.72} &\textbf{-17.67} 
        \\ \cline{2-9}
&\multirow{2}{1cm}{\centering 20 dB}  &Original& -25.23 & -26.02 & -28.24 & -27.96 & -26.99 & -26.78
         \\
       & &Enhanced& \textbf{-25.69} & \textbf{-26.20} & \textbf{-28.54} & \textbf{-28.24} & -26.99 & -26.78
    \\
 \hline
    \end{tabular}}
    \label{tab:MSE_comparison}
\end{table*}

\subsection{MSE performance in high user mobility scenario}

\begin{figure}[h]
\centering
\includegraphics[,width=0.75\linewidth]{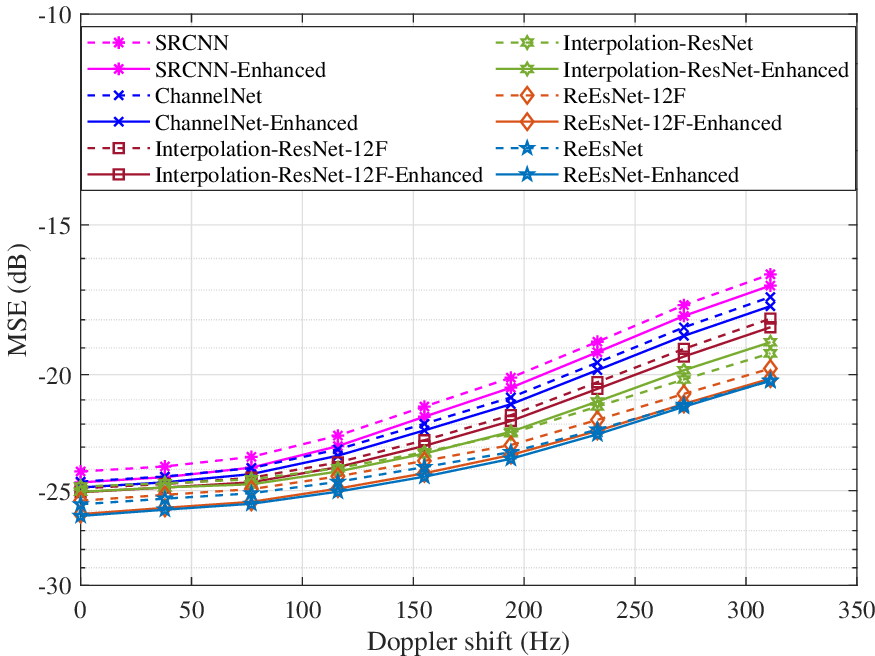}
\caption{MSE versus Doppler shift results comparison of the enhanced and original DL-based channel estimation models for EVA channel model.}
\label{fig:MSE_DOPPLER_EVA}
\end{figure}

In this section, we evaluate the MSE performance of the proposed approach over the Doppler frequency shift ranges for the EVA channel model with a fixed SNR value of 15 dB. We consider mobile speed ranges from $0$ $km/h$ to $160 $ $km/h$ (with an interval of $20$ $km/h$) corresponding to a Doppler frequency shift from 0 Hz to 311 Hz, respectively. 
The MSE performance result for a range of Doppler shift values is shown in Fig.\ref{fig:MSE_DOPPLER_EVA}, where all enhanced models outperform their corresponding original models, with the exception of Interpolation-ResNet-Enhanced at the higher Doppler shift ranges. The non-learnable nature of the upsampling method combined with the Doppler effect limits the inference performance of the Interpolation-ResNet-Enhanced at the higher Doppler shift ranges. Interestingly, ReEsNet-Enhanced and ReEsNet-12F-Enhanced models exhibit significant performance improvements over the original counterparts, particularly in the lower Doppler shift ranges. Overall, the results demonstrate the robustness of the proposed approach against the combined effects of channel variations and high user mobility.

\subsection{Complexity Analysis}

\begin{table*}
  \caption{computational complexity comparison of channel estimation methods.}
   \centering
   \scalebox{0.9}{
    \begin{tabular}{|c |c| c |c|} 
    \hline
        \textbf{Model} & \textbf{Complexity}  \\
        \hline
    LS         & $\mathcal{O}(N_fN_s)$  \\
     \hline 
    MMSE   & $\mathcal{O}(N_f N_s(N_{pf} N_{ps})^2 + (N_{pf} N_{ps})^3)$ \\
          \hline
    
    \makecell{SRCNN \\  SRCNN-Enhanced} &  \makecell{$\mathcal{O}(2N_{f} N_{s} (2\cdot 9^2\cdot64 + 64\cdot1\cdot32 + 32\cdot 5^2 \cdot2))$ \\ 
    $\mathcal{O}(2 N_{f} N_{s} (\mathbf{3}\cdot 9^2\cdot64 +  64\cdot1\cdot32 + 32\cdot 5^2 \cdot2)) $}\\
          \hline
    
    \makecell{ChannelNet \\ ChannelNet-Enhanced} &  \makecell{$\mathcal{O}( \text{SRCNN} + 2N_{f} N_{s} (2 \cdot(2\cdot 3^2\cdot64) + 18 \cdot (64 \cdot 3^2 \cdot 64)))$ \\
    $\mathcal{O}( \text{SRCNN-Enhanced} + 2N_{f} N_{s} (2 \cdot(2\cdot 3^2\cdot64) + 18 \cdot (64 \cdot 3^2 \cdot 64)))$}\\
         \hline
         
    \makecell{ReEsNet \\ ReEsNet-Enhanced}    & 
    \makecell{$\mathcal{O}(2 N_{pf}N_{ps}(3^2 \cdot 2\cdot16 + 8\cdot3^2\cdot16^2 + 3^2\cdot16^2) +  2N_f N_s (11^2\cdot16^2 + 2\cdot3^2\cdot16) )$  \\
    $\mathcal{O}(2 N_{pf}N_{ps}(3^2 \cdot \mathbf{3}\cdot16 + 8\cdot3^2\cdot16^2 + 3^2\cdot16^2) +  2N_f N_s (11^2\cdot16^2 + 2\cdot3^2\cdot16) )$} \\
        \hline
     \makecell{ReEsNet-12F \\ ReEsNet-12F-Enhanced}    & 
    \makecell{$\mathcal{O}(2 N_{pf}N_{ps}(3^2 \cdot 2\cdot \mathbf{12} + 8\cdot3^2\cdot \mathbf{12}^2 + 3^2\cdot \mathbf{12}^2) +  2N_f N_s (11^2\cdot \mathbf{12}^2 + 2\cdot3^2\cdot \mathbf{12}) )$  \\
    $\mathcal{O}(2 N_{pf}N_{ps}(3^2 \cdot \mathbf{3}\cdot \mathbf{12} + 8\cdot3^2\cdot \mathbf{12}^2 + 3^2\cdot \mathbf{12}^2) +  2N_f N_s (11^2\cdot \mathbf{12}^2 + 2\cdot3^2\cdot \mathbf{12}) )$} \\
        \hline
        
    \makecell{Interpolation-ResNet \\ Interpolation-ResNet-Enhanced} & \makecell{$\mathcal{O}(2N_{pf}N_{ps}(3^2 \cdot 2\cdot16 + 8\cdot3^2\cdot16^2 + 3^2\cdot16^2) +  2N_f N_s (2^2\cdot16\cdot16 + 36\cdot7\cdot16\cdot2) )$     \\ $\mathcal{O}(2N_{pf}N_{ps}(3^2 \cdot \mathbf{3}\cdot16 + 8\cdot3^2\cdot16^2 + 3^2\cdot16^2) +  2N_f N_s (2^2\cdot16\cdot16 + 36\cdot7\cdot16\cdot2) )$}\\
        \hline

   \makecell{Interpolation-ResNet-12F \\ Interpolation-ResNet-12F-Enhanced} & \makecell{$\mathcal{O}(2N_{pf}N_{ps}(3^2 \cdot 2\cdot \mathbf{12} + 8\cdot3^2\cdot \mathbf{12}^2 + 3^2\cdot \mathbf{12}^2) +  2N_f N_s (2^2\cdot \mathbf{12}\cdot \mathbf{12} + 36\cdot7\cdot \mathbf{12}\cdot2) )$     \\ $\mathcal{O}(2N_{pf}N_{ps}(3^2 \cdot \mathbf{3}\cdot \mathbf{12} + 8\cdot3^2\cdot \mathbf{12}^2 + 3^2\cdot \mathbf{12}^2) +  2N_f N_s (2^2\cdot \mathbf{12}\cdot \mathbf{12} + 36\cdot7\cdot \mathbf{12}\cdot2) )$}\\
        \hline
    
        \end{tabular}}
    \label{tab:BigOcomplexity}
\end{table*}

The computational complexity and the training overhead comparison of the enhanced and the corresponding original DL-based channel estimation methods are presented in Table \ref{tab:BigOcomplexity} and Table \ref{tab:complexity}, respectively. It can be noted from Table \ref{tab:BigOcomplexity} that the complexity of MMSE increases cubically while that of the DL-based methods increases linearly as the channel dimensions increase. The complexity of ReEsNet-12F and Interpolation-ResNet-12F can be easily obtained by substituting the number of filters of the convolutional layers of ReEsNet and Interpolation-ResNet with 12. Notably, the enhanced DL-based models, particularly the residual learning-based methods, only incur a minor increase in complexity because the adjustments in the input dimensions are made before the upsampling stage.
Further, it can be observed from the training overhead comparison in Table \ref{tab:complexity} that there is no significant increase in the number of model parameters, training time and testing time. The SRCNN-Enhanced and ChannelNet-Enhanced models only add extra 5K parameters. However, both enhanced methods outperform the benchmark counterparts in terms of estimation accuracy. For the remaining methods, there is a negligible increase in the number of model parameters and inference time.  The proposed method has minimal training overhead and thus can be easily deployed for time-constrained applications. 

 \begin{table}
  \caption{DL-based channel estimation Models' training overhead comparison on EPA channel model.}
   \centering
   \scalebox{0.75}{
    \begin{tabular}{c c c c} 
    \hline
        \textbf{Model} & \textbf{Parameters} & \makecell{\textbf{Training time} \\ \textbf{(seconds/epoch)}} &	\makecell{\textbf{Inference time} \\ \textbf{(milliseconds/sample)}} \\
        \hline
    SRCNN & 14,114 & 2 & 0.173 \\
    SRCNN-Enhanced & 19,298 & 2 & 0.227 \\
        \hline 
    ChannelNet & 685,219 & 30 & 0.792 \\
    ChannelNet-Enhanced & 690,403 & 30 & 0.856 \\
         \hline
    ReEsNet    & 52,466  & 9  & 6.67  \\
    ReEsNet-Enhanced    & 52,610  & 9  & 6.77  \\
         \hline 
    ReEsNet-12F  & 29,654  & 9  & 6.30  \\
    ReEsNet-12F-Enhanced  & 29,762  & 9  & 6.30  \\
        \hline
    Interpolation-ResNet & 29,250 & 22 & 9.85 \\
    Interpolation-ResNet-Enhanced & 29,394 & 22 & 10.05 \\
        \hline
    Interpolation-ResNet-12F & 18,050 & 19 & 9.46 \\
    Interpolation-ResNet-12F-Enhanced & 18,158 & 19 & 9.57 \\
        \hline
        \end{tabular}}
    \label{tab:complexity}
\end{table}

\section{CONCLUSION}\label{section5}

This paper proposes a DL-based channel estimation approach with a novel data enhancement method. The proposed data enhancement network, AE-DENet, consists of an AE network that processes the separate real and imaginary data streams of the initial complex LS estimation and extracts the interaction features from the encoder output. These interaction features are then concatenated with the original real and imaginary features to create enhanced LS input, which can be exploited by most existing DL-based channel estimation models to improve the estimation accuracy. The MSE over SNR and Doppler shift results demonstrate that the channel estimation methods with AE-DENet enhanced input outperform the corresponding benchmark methods while adding negligible complexity. The proposed method can effectively utilize the interaction features to improve channel inference and is robust to channel variations and high user mobility.

\bibliographystyle{IEEEtran}
\bibliography{References1.bib}

\begin{thebibliography}{10}
\providecommand{\url}[1]{#1}
\csname url@samestyle\endcsname
\providecommand{\newblock}{\relax}
\providecommand{\bibinfo}[2]{#2}
\providecommand{\BIBentrySTDinterwordspacing}{\spaceskip=0pt\relax}
\providecommand{\BIBentryALTinterwordstretchfactor}{4}
\providecommand{\BIBentryALTinterwordspacing}{\spaceskip=\fontdimen2\font plus
\BIBentryALTinterwordstretchfactor\fontdimen3\font minus \fontdimen4\font\relax}
\providecommand{\BIBforeignlanguage}[2]{{%
\expandafter\ifx\csname l@#1\endcsname\relax
\typeout{** WARNING: IEEEtran.bst: No hyphenation pattern has been}%
\typeout{** loaded for the language `#1'. Using the pattern for}%
\typeout{** the default language instead.}%
\else
\language=\csname l@#1\endcsname
\fi
#2}}
\providecommand{\BIBdecl}{\relax}
\BIBdecl

\bibitem{Ozpoyraz}
B.~Ozpoyraz, A.~T. Dogukan, Y.~Gevez, U.~Altun, and E.~Basar, ``Deep learning-aided {6G} wireless networks: A comprehensive survey of revolutionary {PHY} architectures,'' \emph{IEEE Open J. Commun. Soc.}, vol.~3, pp. 1749--1809, Sept. 2022.

\bibitem{9128890}
A.~K. Gizzini, M.~Chafii, A.~Nimr, and G.~Fettweis, ``{Enhancing Least Square Channel Estimation Using Deep Learning},'' in \emph{2020 IEEE 91st Veh. Technol. Conf. (VTC2020-Spring)}, May 2020, pp. 1--5.

\bibitem{8052521}
H.~Ye, G.~Y. Li, and B.-H. Juang, ``{Power of Deep Learning for Channel Estimation and Signal Detection in OFDM Systems},'' \emph{IEEE Wireless Commun. Lett.}, vol.~7, no.~1, pp. 114--117, Feb. 2018.

\bibitem{8715338}
H.~He, S.~Jin, C.-K. Wen, F.~Gao, G.~Y. Li, and Z.~Xu, ``{Model-Driven Deep Learning for Physical Layer Communications},'' \emph{IEEE Wireless Commun.}, vol.~26, no.~5, pp. 77--83, Oct. 2019.

\bibitem{8509622}
X.~Gao, S.~Jin, C.-K. Wen, and G.~Y. Li, ``{ComNet: Combination of Deep Learning and Expert Knowledge in OFDM Receivers},'' \emph{IEEE Commun. Lett.}, vol.~22, no.~12, pp. 2627--2630, Dec. 2018.

\bibitem{8761312}
Y.~Liao, Y.~Hua, X.~Dai, H.~Yao, and X.~Yang, ``{ChanEstNet: A Deep Learning Based Channel Estimation for High-Speed Scenarios},'' in \emph{ICC 2019 - 2019 IEEE Int. Conf. Commun. (ICC)}, May 2019, pp. 1--6.

\bibitem{8640815}
M.~Soltani, V.~Pourahmadi, A.~Mirzaei, and H.~Sheikhzadeh, ``{Deep Learning-Based Channel Estimation},'' \emph{IEEE Commun. Lett.}, vol.~23, no.~4, pp. 652--655, Apr. 2019.

\bibitem{9815290}
W.~Tong, W.~Xu, F.~Wang, J.~Shang, M.~Pan, and J.~Lin, ``{Deep Learning Compressed Sensing-Based Beamspace Channel Estimation in mmWave Massive MIMO Systems},'' \emph{IEEE Wireless Commun. Lett.}, vol.~11, no.~9, pp. 1935--1939, Sept. 2022.

\bibitem{9120030}
A.~K. Gizzini, M.~Chafii, A.~Nimr, and G.~Fettweis, ``{Deep Learning Based Channel Estimation Schemes for IEEE 802.11p Standard},'' \emph{IEEE Access}, vol.~8, pp. 113\,751--113\,765, Jun. 2020.

\bibitem{8933411}
Y.~Liao, Y.~Hua, and Y.~Cai, ``{Deep Learning Based Channel Estimation Algorithm for Fast Time-Varying MIMO-OFDM Systems},'' \emph{IEEE Commun. Lett.}, vol.~24, no.~3, pp. 572--576, Mar. 2020.

\bibitem{10052744}
F.~Zhang, C.~Luo, J.~Xu, and Y.~Luo, ``{An Autoencoder-based I/Q Channel Interaction Enhancement Method for Automatic Modulation Recognition},'' \emph{IEEE Trans. Veh. Technol.}, pp. 1--6, Feb. 2023.

\bibitem{10051858}
M.~Baur, B.~Fesl, M.~Koller, and W.~Utschick, ``{Variational Autoencoder Leveraged MMSE Channel Estimation},'' in \emph{2022 56th Asilomar Conf. Signals, Syst. Comput.}, Nov. 2022, pp. 527--532.

\bibitem{9166541}
M.~Soltani, V.~Pourahmadi, and H.~Sheikhzadeh, ``{Pilot Pattern Design for Deep Learning-Based Channel Estimation in OFDM Systems},'' \emph{IEEE Wireless Commun. Lett.}, vol.~9, no.~12, pp. 2173--2176, Dec. 2020.

\bibitem{chen}
H.-Y. Chen, M.-H. Wu, T.-W. Yang, C.-W. Huang, and C.-F. Chou, ``Attention-aided autoencoder-based channel prediction for intelligent reflecting surface-assisted millimeter-wave communications,'' \emph{IEEE Trans. Green Commun. Netw.}, vol.~7, no.~4, pp. 1906--1919, May 2023.

\bibitem{li2019deep}
L.~Li, H.~Chen, H.-H. Chang, and L.~Liu, ``Deep residual learning meets {OFDM} channel estimation,'' \emph{IEEE Wireless Commun. Lett.}, vol.~9, no.~5, pp. 615--618, 2019.

\bibitem{luan2021low}
D.~Luan and J.~Thompson, ``Low complexity channel estimation with neural network solutions,'' in \emph{WSA 2021; 25th Int. ITG Workshop Smart Antennas}.\hskip 1em plus 0.5em minus 0.4em\relax VDE, 2021, pp. 1--6.

\end{thebibliography}

\end{document}